\documentstyle[psfig]{mn}
\newif\ifAMStwofonts

\ifoldfss
  \ifCUPmtlplainloaded \else
    \NewTextAlphabet{textbfit} {cmbxti10} {}
    \NewTextAlphabet{textbfss} {cmssbx10} {}
    \NewMathAlphabet{mathbfit} {cmbxti10} {} % for math mode
    \NewMathAlphabet{mathbfss} {cmssbx10} {} %  "   "    "
  \fi
  \ifAMStwofonts
    \ifCUPmtlplainloaded \else
      \NewSymbolFont{upmath} {eurm10}
      \NewSymbolFont{AMSa} {msam10}
      \NewMathSymbol{\upi}     {0}{upmath}{19}
      \NewMathSymbol{\umu}     {0}{upmath}{16}
      \NewMathSymbol{\upartial}{0}{upmath}{40}
      \NewMathSymbol{\leqslant}{3}{AMSa}{36}
      \NewMathSymbol{\geqslant}{3}{AMSa}{3E}

      \let\leq=\leqslant 
      \let\geq=\geqslant 
    \fi
  \fi
\fi % End of OFSS

\ifnfssone
  \newmathalphabet{\mathit}
  \addtoversion{normal}{\mathit}{cmr}{m}{it}
  \addtoversion{bold}{\mathit}{cmr}{bx}{it}
  \newmathalphabet{\mathbfit} % math mode version of \textbfit{..}
  \addtoversion{normal}{\mathbfit}{cmr}{bx}{it}
  \addtoversion{bold}{\mathbfit}{cmr}{bx}{it}
  \newmathalphabet{\mathbfss} % math mode version of \textbfss{..}
  \addtoversion{normal}{\mathbfss}{cmss}{bx}{n}
  \addtoversion{bold}{\mathbfss}{cmss}{bx}{n}
  \ifAMStwofonts
    \ifCUPmtlplainloaded \else
      \UseAMStwoboldmath
      \makeatletter
      \new@mathgroup\upmath@group
      \define@mathgroup\mv@normal\upmath@group{eur}{m}{n}
      \define@mathgroup\mv@bold\upmath@group{eur}{b}{n}
      \edef\UPM{\hexnumber\upmath@group}
      \new@mathgroup\amsa@group
      \define@mathgroup\mv@normal\amsa@group{msa}{m}{n}
      \define@mathgroup\mv@bold\amsa@group{msa}{m}{n}
      \edef\AMSa{\hexnumber\amsa@group}
      \makeatother
      \mathchardef\upi="0\UPM19
      \mathchardef\umu="0\UPM16
      \mathchardef\upartial="0\UPM40
      \mathchardef\leqslant="3\AMSa36
      \mathchardef\geqslant="3\AMSa3E

      \let\leq=\leqslant 
      \let\geq=\geqslant 
    \fi
  \fi
\fi % End of NFSS release 1

\ifnfsstwo
  \DeclareMathAlphabet{\mathbfit}{OT1}{cmr}{bx}{it}
  \SetMathAlphabet\mathbfit{bold}{OT1}{cmr}{bx}{it}
  \DeclareMathAlphabet{\mathbfss}{OT1}{cmss}{bx}{n}
  \SetMathAlphabet\mathbfss{bold}{OT1}{cmss}{bx}{n}
  \ifAMStwofonts
    \ifCUPmtlplainloaded \else
      \DeclareSymbolFont{UPM}{U}{eur}{m}{n}
      \SetSymbolFont{UPM}{bold}{U}{eur}{b}{n}
      \DeclareSymbolFont{AMSa}{U}{msa}{m}{n}
      \DeclareMathSymbol{\upi}{0}{UPM}{"19}
      \DeclareMathSymbol{\umu}{0}{UPM}{"16}
      \DeclareMathSymbol{\upartial}{0}{UPM}{"40}
      \DeclareMathSymbol{\leqslant}{3}{AMSa}{"36}
      \DeclareMathSymbol{\geqslant}{3}{AMSa}{"3E}

      \let\leq=\leqslant 
      \let\geq=\geqslant 
    \fi
  \fi
\fi % End of NFSS release 2

\ifCUPmtlplainloaded \else
  \ifAMStwofonts \else % If no AMS fonts
    \def\upi{\pi}
    \def\umu{\mu}
    \def\upartial{\partial}
  \fi
\fi

\title{Quasar--galaxy associations revisited}

\author[Ben\'\i tez, Sanz \& Mart\'\i nez-Gonz\'alez]{
N. Ben\'{\i}tez$^{1,2}$, J. L. Sanz$^3$ \& E. Mart\'\i nez-Gonz\'alez$^3$\\
$^1$ Astronomy Department, UC Berkeley, 601 Campbell Hall, Berkeley CA 94704\\
$^2$ Dept. of Physics and Astronomy, JHU, 3400 N.Charles St., Baltimore MD 21218\\
$^3$ IFCA (CSIC-UC), Avenida de Los Castros s/n 39001, Santander, SPAIN}

\date{}

\begin{document}

\maketitle

\begin{abstract}
Gravitational lensing predicts an enhancement of the density of bright, distant 
QSOs around foreground galaxies. We measure this QSO--galaxy correlation \( w_{qg} \)
for two \textit{complete} samples of radio-loud quasars, the southern 1Jy and
Half-Jansky samples. The existence of a positive correlation between 
\( z\sim 1 \) quasars and \( z\approx 0.15 \) galaxies is confirmed at a \( p=99.0\% \)
significance level (\( >99.9\% \) if previous measurements on the northern hemisphere 
are included). A comparison with the results obtained for incomplete quasar catalogs 
(e.g. the Veron-Cetty and Veron compilation) suggests the existence of an 
`identification bias', which spuriously increases the estimated amplitude 
of \( w _{qg} \) for incomplete samples. This effect may explain many of 
the surprisingly strong quasar--galaxy associations found in the literature. 
Nevertheless, the value of \( w_{qg} \) that we measure in our complete catalogs 
is still considerably higher than the predictions from weak lensing. We consider 
two effects which could help to explain this discrepancy: galactic dust extinction 
and strong lensing. 

\end{abstract}

\begin{keywords}
quasar galaxy associations; gravitational lensing
\end{keywords}

\section{Introduction}

Canizares (1981) showed that if gravitational lensing effects are considered,
an enhancement in the density of QSOs close to the position of foreground galaxies
is expected (see also Keel 1982; Peacock 1982). This effect, known as `magnification
bias' can be characterized by the overdensity \( q=\mu ^{\alpha -1} \) (Narayan
1989), where \( \mu  \) is the magnification and \( \alpha  \) is the logarithmic
slope of the number counts distribution. Depending on the
slope \( \alpha  \), an excess (\( q>1 \)) or even a defect (\( q<1 \)) of
background sources around the lenses will be observed. Galaxies and clusters
trace the matter fluctuations (up to a bias factor \( b \)), and therefore,
we may observe positive (\( \alpha >1 \)), null (\( \alpha =1 \)) or negative
(\( \alpha <1 \)) statistical associations of these foreground objects with
distant, background sources.

The evidence for association between high--redshift AGNs and foreground galaxies
has steadily accumulated during the years (see Schneider, Ehlers \& Falco 1992 
for a detailed review). However, the results are sometimes apparently contradictory 
and quite often difficult to accommodate within the gravitational lensing framework. 
For instance, the strength of associations depends on the AGN type and in particular, 
the studies performed with radio-loud AGN samples, or with heterogeneous samples, 
usually extracted from early versions of the Veron-Cetty \& Veron or the Hewitt \& Burbidge 
catalogs, (which contain a high proportion of radio--loud QSOs) almost routinely find 
significant excesses of foreground objects around the quasar positions (Tyson 1986; 
Fugmann 1988,1990; Hammer \& Le F\'{e}vre 1990; Hintzen et al. 1991; Drinkwater et al. 1992;
Thomas et al. 1995; Bartelmann \& Schneider 1993b, 1994; Bartsch, Schneider,
\& Bartelmann 1997; Seitz \& Schneider; Ben\'{\i}tez et al. 1995; Ben\'{\i}tez
\& Mart\'{\i}nez-Gonz\'{a}lez 1995, 1997(BMG97); 
Ben\'{\i}tez, Mart\'{\i}nez-Gonz\'{a}lez \& Mart\'{\i}n-Mirones 1997; 
Norman \& Williams 1999; Norman \& Impey 1999). 
Although these results are qualitatively in agreement with the magnification 
bias effect, in most cases the amplitude of the correlation is much higher 
than that expected from gravitational lensing models.

On the other hand, the studies carried out with optically selected catalogs
are less consistent. Positive correlations have been found by Webster et al. 
(1988), Rodrigues-Williams and Hogan(1991) and Williams \& Irwin (1998). In all 
these cases, the amplitude of the correlations is more than an order of magnitude
stronger than the predictions from the magnification bias effect. On the other
hand, Boyle et al. (1988); Romani \& Maoz (1992); BMG97; 
Ferreras, Ben\'{\i}tez \& Mart\'{\i}nez-Gonz\'{a}lez (1997) and 
Croom \& Shanks (1999) found null or even negative correlations, which in some 
cases are expected from the magnification bias effect due to the shallowness of the 
QSO number counts. However, there are certain instances in which the differences 
in the QSO--galaxy correlation $w_{qg}$ between radio--loud and radio--quiet QSOs 
cannot be explained by the lensing hypothesis: BMG97 found a negative correlation 
for the optically selected Large Bright Quasar Survey (LBQS) catalog 
(Hewett, Foltz \& Chafee 1995), which has a steep slope $\alpha \gse 2$. 
The most likely explanation for this is that the LBQS catalog is 
affected by a selection bias which hinders the detection of high-z QSOs 
in regions of high projected galaxy density (see Romani \& Maoz 1992, Maoz 1995). 
Further evidence was obtained by Ferreras, Ben\'{\i}tez \& 
Mart\'{\i}nez-Gonz\'{a}lez (1997) which found that CFHT/MMT QSOs 
(Crampton, Cowley \& Hartwick 1989) were strongly anticorrelated with low redshift
galaxies, contrary to what was expected from the magnification bias effect.
The amplitude of the anticorrelation is stronger for lower redshift QSOs which
practically excludes dust in foreground galaxies as its cause and strongly suggests 
a dependence on the strength of the emission lines used to identify the QSO.
From these results it may be concluded that until the selection effects operating
on optically selected QSO samples are not better understood, one should be wary
of the gravitational lensing inferences obtained from them.

As Bartelmann \& Schneider (1993a) pointed out, the scale of some of the positive
correlations reported above (several arcmin) is difficult to explain considering
lensing by isolated galaxies or microlensing, and apparently has to be caused
by the large-scale structure. Bartelmann (1995) showed that in the weak lensing regime 
\begin{equation}
w _{qg}(\theta )=b(\alpha -1)C_{\mu \delta }
\label{uno}
\end{equation}
 where \( C_{\mu \delta } \) is the correlation between magnification and matter
density contrast \( C_{\mu \delta }=<\delta \mu \delta _{m}> \), \( b \) is
the biasing factor \( b=\delta _{g}/\delta _{m} \), where \( \delta _{g} \)
and \( \delta _{m} \) are respectively the galaxy and dark matter overdensities.
Sanz et al. (1997) and Dolag \& Bartelmann (1997) have calculated \( C_{\mu \delta } \)
for several cosmological models taking into account the non linear evolution
of the power spectrum of density perturbations. In Ben\'\i tez \& Sanz (1999) it was
shown that the expected value of \( w_{qg} \) can also be estimated as
\begin{equation}
w_{qg}=Q(z_{b},z_{f})\Omega_m b^{-1}w_{gg}
\label{dos}
\end{equation}
where $z_f$ and $z_b$ are the typical redshifts of respectively the galaxy and quasar 
samples. For low redshift galaxies and relatively high-z QSOs like the ones considered
in this paper, the factor \( Q \) is approximately independent of the cosmological
model.

Any radio--loud quasar is detected in the optical, with e.g. magnitude \( B \), 
and in radio with flux \( S \). For a sample with independent optical and radio fluxes, 
the effective slope \( \alpha _{e} \) (Borgeest et al. 1991) takes the form 
\begin{equation}
\alpha _{e}(S,B)=-{d\ln N(>S)\over d\ln S}+{2.5d\log N(<B)\over dB}
\end{equation}
 where \( N(>S) \) and \( N(<B) \) are the cumulative number counts distributions. 
This is known as the double magnification bias effect. It holds if the optical and 
radio fluxes are independent, and also if the source sizes are considerable smaller than 
the lens in both bands, what ensures that both fluxes are equally magnified. Both 
conditions are reasonably fulfilled for QSO--galaxy associations: radio and optical 
fluxes are practically uncorrelated for radio quasars (see e.g. Fig. 10 of Drinkwater et al. 1997) 
and the large scale structure weak lensing field is smooth enough to magnify similarly the 
sizes of the optical and radio emitting regions. 

Therefore, to compare the model predictions with the observations, it is necessary
to know in detail the unperturbed number counts distribution of the quasar sample.
However, practically all the radio--loud samples mentioned above are incomplete
(e.g. BMG97) or have heterogeneous photometrical information in the optical 
(e.g. Ben\'\i tez \& Mart\'\i nez-Gonz\'alez 1995). 
Here we intend to remedy this situation by measuring \( w_{qg} \) using 
two {\it complete} radioloud quasar samples.

The outline of the paper is the following. In \S2 we describe the galaxy 
and QSO samples used to estimate \( w _{qg} \). \S3 deals with the 
statistical analysis. \S4 presents two theoretical estimations of $w_{qg}$ 
and compares them with the observational results. \S5 discusses the puzzling 
results of the previous section and \S6 summarizes our main results and conclusions. 

\section{The Data}

\label{data}

Our galaxy sample is obtained from the ROE/NRL COSMOS/UKST Southern Sky catalog
(see Yentis et al. 1992 and references therein), which contains the objects detected
in the COSMOS scans of the glass copies of the ESO/SERC blue survey plates.
The UKST survey is organized in \( 6\times 6\deg ^{2} \) fields on a \( 5\deg  \)
grid and covers the zone with declination \( -90\deg <\delta <0\deg  \) and
galactic latitude \( |b|>10\deg  \). The catalog supplies several parameters
for each detected object, including the centroid in both sky and plate coordinates,
\( B_{J} \) magnitude and the star-galaxy image classification down to a limiting
magnitude of \( B_{J}\approx21  \). Drinkwater et al. (1997) quote a calibration
accuracy of about \( \pm 0.5 \)mag for the COSMOS \( B_{J} \) magnitudes.
As in BMG97, we include in our sample galaxies brighter than \( B_{J}<20.5 \)
and within \( 0.2\arcmin <\theta <15\arcmin  \) of the quasars selected as
we describe below. The median redshift of the \( B_{J} \) galaxies is 
\( z_{g}\approx 0.15 \).

In BMG97 we studied the quasar-galaxy correlation for a sample of 144 \( z>0.3 \)
PKS quasars with a 2.7GHz flux \( S_{2.7}>0.5\)Jy, extracted from Veron-Cetty \& 
Veron (1996). This sample was shown to be 
strongly correlated with foreground
galaxies, but its incompleteness precluded a detailed comparison with theory.
Thus, we now consider two well defined and practically complete radio loud quasar
catalogs to assess the incompleteness of the BMG97 sample and its effect on the
estimation of \( w _{qg} \): the 1Jy catalog (Stickel, 
Meisenheimer \& K\"{u}hr 1994; Stickel et al. 1996)
which includes radiosources with a 5GHz flux \( S_{5}>1 \)Jy, and the Parkes
Half-Jansky Flat-Spectrum sample (Drinkwater et al. 1997) which contains flat-spectrum
sources detected at 2.7GHz and 5GHz, with S\( _{2.7}\geq 0.5 \)Jy, spectral
index between 2.7GHz and 5GHz \( \alpha _{2.7/5.0}>-0.5 \), galactic latitude
\( |b|>20\deg  \) and declination \( -45\deg <\delta <10\deg  \).

Both the 1Jy and Half-Jansky catalogs are almost \( 100\% \) identified and
have spectroscopical redshifts for \( \geq 90\% \) of the sources. 
To try to work with samples as similar as possible, we have extracted from the 
three catalogs those objects classified as QSOs. VC classify as QSOs those objects 
with absolute magnitudes brighter than $B=-23$ and the 1Jy catalog uses the same
classifications as the original discovery papers. For the HJ we have selected 
those objects classified as ``stellar'', excluding the BL LACs to get a more homogeneous 
sample. Drinkwater \& Schmidt 1996 showed that the original Parkes catalog classification 
was made in a non-uniform way, resulting in false large-scale correlations 
among quasars drawn from that catalog. The HJ classification seems to be free of 
this problem (Drinkwater et al. 1997) and even if the 1Jy presents a similar defect, 
it would not affect the measurement of $w_{qg}$, since we normalize the galaxy 
density locally in a 15' circle around each quasar. 
It should also be kept in mind that the optical classification of radiosources 
may be affected by crowding effects (an example are the 'merged' objects in Drinkwater et 
al. 1997), and that this could tend to eliminate from the sample QSOs in 
fields with higher than average galaxy density, biasing low the 
measured value of $w_{qg}$. 

Our analysis will be performed within basically the same area
defined by Drinkwater et al. (1997), except for three additional constraints imposed
by the characteristics of the COSMOS/UKST galaxy catalog:

i) Declinations \( -45\deg <\delta <3\deg  \), as the COSMOS/UKST plates only
reach up to \( \delta <3\deg  \).

ii) Galactic latitude \( |b|>30\deg  \). It was shown in BMG97 that the ``galaxy''
density in COSMOS fields is significantly anticorrelated with galaxy latitude,
clearly due to an increase in the numbers of stars misclassified as ``galaxies''
when we approach the galactic disk. In fact, COSMOS fields with \( |b|<30\deg  \)
have on average \( 75\% \) more objects classified as ``galaxies'' than the
\( |b|>30\deg  \) fields. Although excluding these fields does not affect
strongly the results (BMG97), they would dilute the galaxy excess and bias low
the amplitude of \( w _{qg} \).

iii) \( |\Delta x| \),\( |\Delta y|<2.25\deg  \), 
\( \sqrt{\Delta x^{2}+\Delta y^{2}}<2.5\deg  \).
As in BMG97, these cutoffs avoid the outer regions of the plates, with worse
image and photometric qualities. We also exclude a few fields from the sample
because they are affected by meteorite traces.

There are also other three general constraints based on the quasar characteristics:

iv) A faint threshold on the quasar \( B_{J} \) magnitude, \( B_{J}<20.5 \),
the same as for our galaxies. We also set an upper threshold of \( B_{J}>15 \),
brighter of which the photographic magnitudes are not reliable. We use only
COSMOS \( B_{J} \) magnitudes obtained from Drinkwater et al. (1997), or directly
from the ROE/NRL COSMOS/UKST Southern Sky catalogue. This ensures photometrical
bandpass homogeneity. At the chosen magnitude limit, it can be reasonably expected
that any radio source candidate is bright enough to be detected, correctly identified
on a photographic plate and have its redshift determined spectroscopically.
Up to this limit, the quasar catalogs should be practically complete.

v)A lower redshift cutoff, \( z>0.3 \). Only \( 5\% \) of the COSMOS/UKST
galaxies with \( B_{J}<20.5 \) have \( z>0.3 \), so we exclude the possibility
of ``intrinsic'' quasar-galaxy correlations contaminating the result. Setting
a higher redshift cutoff does not affect significantly the results.

vi)A 2.7GHz flux cutoff: \( S_{2.7}\geq 0.5 \) Jy. This is the flux limit of
the Half-Jansky catalog. The \( S_{2.7} \) fluxes are obtained from Drinkwater
et al. (1997) and from the Veron-Cetty \& Veron compilation. Objects from the
1Jy catalog also have \( S_{5}>1 \) Jy. We do not set any constrains on the
steepness of the radio spectra for the Veron-Cetty \& Veron and 1Jy samples.
The Half-Jansky sample has \( \alpha _{2.7/5.0}>-0.5 \).

\begin{figure}
%\vspace{8.89cm}
\psfig{figure=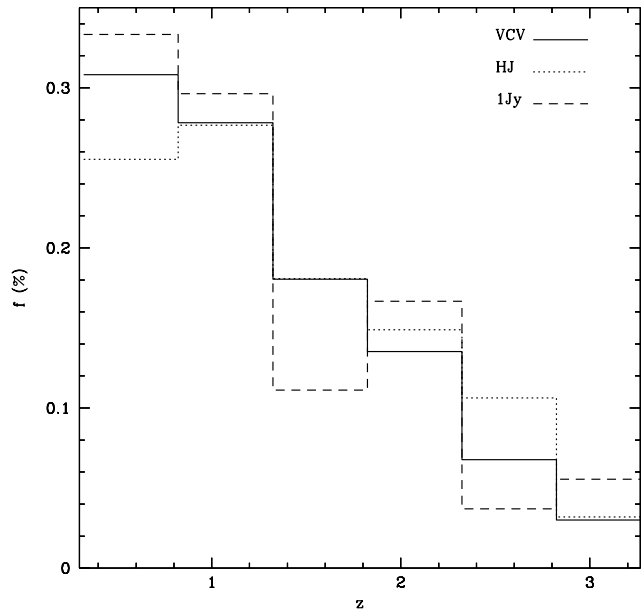,height=3.5in}
\caption{Redshift histograms for the VC (solid line), HJ (dashed
line) and 1Jy (dotted line) samples. The vertical axis show the fraction of
quasars in each redshift bin.} 
\end{figure}

The Veron-Cetty \& Veron (VC) sample is very similar to the PKS sample used
in BMG97 except for the exclusion of 23 quasars with \( |b|<30\deg  \) and
\( B_{J}<20.5 \), and the inclusion of 12 objects which belong to the Veron-Cetty
\& Veron (1997) catalog and comply with criteria i-vi) but were not included
in BMG97 because their names do not start with ``PKS'', or have \( S_{2.7} \)
exactly equal to 0.5Jy (the flux limit in BMG97 was \( S_{2.7}>0.5Jy \) and
here it is \( S_{2.7}\geq 0.5Jy \)). The VC sample thus defined contains \textit{all}
the quasars in the 1Jy sample and \( 91.5\% \) of the Half-Jansky (HJ) sample
(86 out of 94).

There are 35 VC quasars which do not belong to the HJ or 1Jy sample. They are steep-spectrum
quasars fainter than the 1Jy catalog $ S_{5}>1$ Jy  limit, and are slightly brighter
in the optical (\( <B_{J}>=17.50 \)) than the HJ sample. On the other hand,
there are 8 HJ quasars which are not included in the VC sample. They are considerable
fainter than the rest of the HJ sample, with \( B_{J}>18.6 \) and \( S_{2.7}<0.8 \).

\begin{figure}
%\vspace{8.89cm}
\psfig{figure=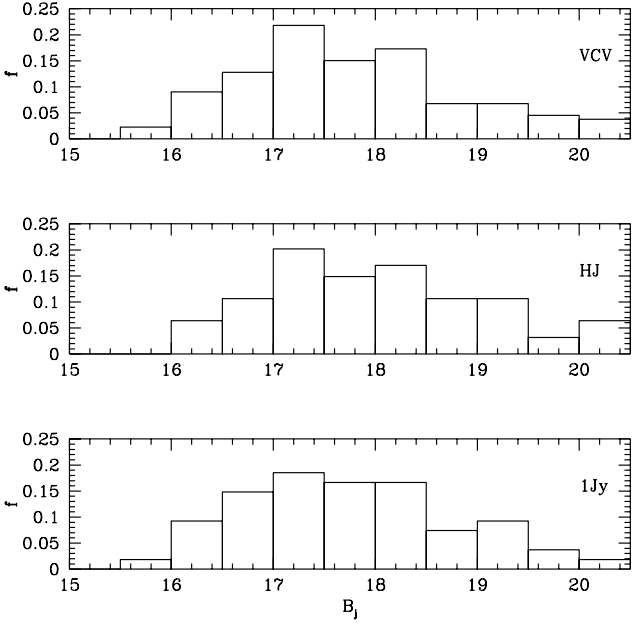,height=3.5in}
\caption{\protect\protect\protect\( B_{J}\protect \protect \protect \)
histograms for the VC, HJ and 1Jy samples. The vertical axis show the fraction
of quasars in each bin.} 
\end{figure}

The redshift distributions of the HJ, VC and 1Jy samples are shown in Fig 1.
The distributions look rather similar, although the HJ quasars apparently tend
to have slightly higher redshifts than the 1Jy ones. However, a Kolmogorov-Smirnov
test shows that the difference between both distributions has only a \( 69\% \)
significance level. As Drinkwater et al. (1997) showed, the redshift distribution
of the HJ catalog is very similar to that of optically selected samples, as
the LBQS (Hewett, Foltz \& Chafee 1995). Fig 2 displays the \( B_{J} \) histograms of the HJ,
VC and 1Jy samples. Although VC and 1Jy quasars are slightly brighter than HJ
ones, the difference is again not statistically significant. The cumulative
number distributions $N(<B_J)$ of the \( z>0.3 \) quasars in the HJ and 1Jy samples  
are shown in Fig. 3. They are much shallower than the corresponding distribution for 
optically-selected QSOs (Hawkins \& Veron 1995). This fact was also noticed
in BMG97 but since we were using a subsample of an incomplete catalog, it was not possible
to distinguish between the effect of incompleteness and the intrinsic differences
in the luminosity functions. The number counts \( N(<B_{J}) \) for the quasars
are well fitted by a law of the form

\[
\log N(<B_{J})=\log N_{B}+a_{1}\Delta B_{J}-a_{2}\Delta B_{J}^{2},\, \, \, \, \, \, \, \, \, \, \, \, \Delta B_{J}\leq 0\]

\[
\log N(<B_{J})=\log N_{B}+a_{1}\Delta B_{J},\, \, \, \, \, \, \, \, \, \, \, \, \, \, \, \, \, \, \, \, \, \, \, \, \Delta B_{J}\geq 0\]
where \( \Delta B_{J}=B_{J}-18.75 \) and \( N_{B} \) is an arbitrary normalization.
For the 1Jy sample \( a_{1}=0.19,\, \, \, \, a_{2}=0.66 \) , whereas for the
HJ sample, \( a_{1}=0.11,\, \, \, \, a_{2}=0.77 \). The above fits are plotted in
Fig. 3.

\begin{figure}
%\vspace{8.89cm}
\psfig{figure=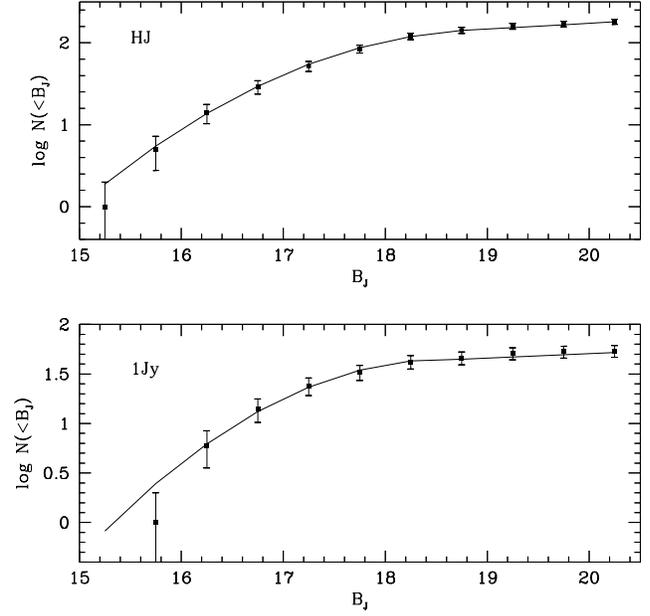,height=3.5in}
\caption{Cumulative number counts--magnitude distribution for
the HJ and 1Jy samples. The solid lines are the least squares fits described in
Sec 2}
\end{figure}

The cumulative flux distribution of the \( z>0.3 \) quasars in the HJ and 1Jy
catalogs are shown in Figure 4. For the HJ sample 
\begin{equation}
\log N(>S_{2.7})=\log N_{0}-1.46\log S_{2.7}-0.36(\log S_{2.7})^{2}
\end{equation}
 and for the 1Jy sample 
\begin{equation}
\log N(>S_{5})=\log N_{0}-1.82\log S_{5}
\end{equation}
 where \( N_{0} \) is an arbitrary normalization. 

\begin{figure}
%\vspace{8.89cm}
\psfig{figure=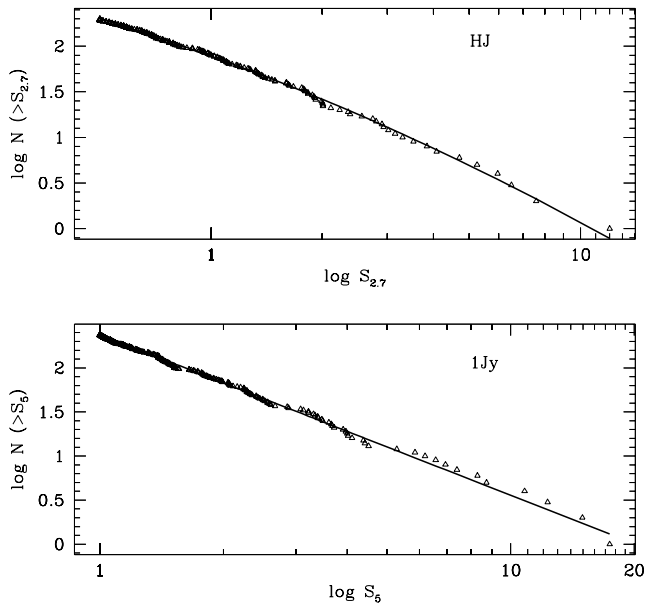,height=3.5in}
\caption{Cumulative number counts--flux distribution for the HJ and 1Jy
catalogs (we include all the \protect\protect\protect\( z>0.3\protect \protect \protect \)
quasars in the original Half-Jansky and 1Jy catalogs). The solid lines are least
squares fits}
\end{figure}

Therefore, at the flux limits of the HJ sample, \( B_{J}=20.5 \) and \( S_{2.7}=0.5 \)Jy,
we have that \( (\alpha _{e}-1)_{HJ}=0.18+1.24-1=0.42 \). For the 1Jy sample,
\( (\alpha _{e}-1)_{1Jy}=0.11+1.82-1=0.93 \). The real value of \( \alpha _{e}-1 \)
is therefore \( 2-4 \) times smaller than the values so far considered in the
estimation of \( w _{qg} \) and \( q \) for the PKS sample (BMG97, Dolag
\& Bartelmann 1997), \( \alpha _{e}-1=1.8-2.2 \).

\section{Statistical Analysis}

\begin{table}
% \centering
% \begin{minipage}{140mm}
\caption{Results of the weighted average test}
\begin{tabular}{lcccc}

Sample & \( N_{Q} \)& \( N_{g} \)& \( r_{w} \)& \( p_w \) \\
VC            &  133       & 12662      & 1.01375    & 99.55\% \\ 
1Jy           &  54        & 5339       & 1.01537    & 97.09\% \\ 
HJ            &  94        & 9243       & 1.01178    & 97.33\% \\ 
HJ + 1Jy      &  106       & 10415      & 1.01200    & 97.84\% \\ 
VC+HJ+1Jy     &  141       & 13577      & 1.01134    & 98.57\% \\ 
VC ``faint''  &  8         & 553        & 1.03805    & 93.43\% \\ 
HJ ''faint''  &  14        & 1309       & 0.99953    & 50.37\% \\ 
``non-VC'' HJ &  8         & 915        & 0.97132    &  6.72\% \\ 
\end{tabular}

\medskip

$N_{Q}$ is the number of QSOs in each sample, $N_{g}$ the number of galaxies 
for which the $w_{qg}$ is measured, $r_{w}$ is the value of the normalized 
correlation coefficient (see text) and $p_w$ the corresponding significance 
associated to the value of $r_w$.

%\end{minipage}
\end{table}

\label{stats}

Spearman's rank order test has often been used to study the statistical 
significance of QSO-galaxy associations. For instance, in the 
implementation of Bartelmann \& Schneider (1993a, 1994), all 
the individual galaxy fields around the quasars 
are merged into a single `superfield', which is subsequently divided into
$N_{bins}$ annular intervals of equal surface. The rank order test 
determines whether the number of galaxies in each bin $n_i,( i=1,N_{bins})$ is 
anticorrelated with the bin radius $r_i$. Here we have applied a 
variant of this test described in detail in BMG97. The field is divided into rings 
of fixed width $\Delta \theta$ and distance $\theta_i$ from the QSO, 
and the variables to be correlated are $w(\theta_i)$, the value of 
the empirical angular correlation in the $i-$bin, and $\theta_i$. The results 
of several binnings are averaged to reduce the dependence of the significance 
on the particular choice of $\Delta\theta$. Extensive Montecarlo simulations show 
the robustness of this approach.

The main advantage of this test is that it does not rely on any particular shape
of the correlation function \( w_{qg} \). It just tells us whether the distribution
of galaxies is correlated with the positions of the QSOs. From it we 
find that the 1Jy sample shows a positive correlation with galaxies at a
\( 99.0\% \) confidence level. Note that the 1Jy 'North', \( \delta >0 \)
sample which is also practically complete and does not overlap with
the sample considered here, was found to be correlated with foreground APM galaxies
at a \( 91.4\% \) level (\( 99.1\% \) for the red galaxy sample 
Ben\'\i tez \& Mart\'\i nez-Gonz\'alez 1995; see also 
Norman \& Williams 1999). A naive but conservative combination of both results yields 
\( p>99.9 \) for the existence of positive correlations between the whole 
1Jy catalog and foreground galaxies. The existence of correlation for the HJ sample
is detected at a smaller significance level, \( p=84.1\% \).

To compare among the different samples we found more convenient the 
weighted average test of Bartsch et al. (1997). This test is based on the
estimator \( r_{g} \), 
\begin{equation}
r_{g}=\frac{1}{N}\sum _{i=1}^{N}w _{qg}(\theta _{i})
\end{equation}
 where \( N \) is the total number of galaxies and \( \theta _{i} \) are the
quasar-galaxy distances. Bartsch et al. (1997) showed that \( r_{g} \) is optimized
to distinguish between a random distribution of galaxies around the quasars
and a distribution which follows an assumed \( w _{qg} \).

We will slightly modify the procedure of Bartsch et al. (1997) and use a normalized
value of \( r_{g} \), which we define as \( r_{w }=r_{g}/I_{w _{qg}} \)
where 
\begin{equation}
I_{w _{qg}}=\frac{2\pi }{N}\int _{\theta _{min}}^{\theta _{max}}w _{qg}(\theta )\theta d\theta 
\end{equation}
Values of $r_w>1$ will indicate a positive correlation, $r_w<1$ negative, and $r_w=0$ 
absence of correlation. If the 
galaxies are exactly distributed following \( w _{qg} \), then
\( r_{g} \) happens to be the Monte-Carlo integral of \( w _{qg}^{2} \),
i.e., \( r_{g} \equiv I_{w _{qg}^{2}} \). Unlike Bartsch et al. (1997) we 
do not merge all the fields into one single superfield, but calculate \( r_{w } \) 
for each quasar and then found an average \( <r_{w }> \) over all the sample. This
avoids giving more weight to a field just because its galaxy density is higher,
which could be due to star contamination.

We set \( w_{qg}\propto \theta ^{-0.688} \) as predicted by Ben\'{\i}tez \&
Sanz (1999), and proportional to the galaxy angular correlation of the APM galaxies
(Maddox et al. 1990). This test is very robust in the sense that it is only sensitive to
changes in the shape of \( w _{qg}(\theta ) \); if we multiply this function by an 
arbitrary amplitude, the significance will not change. The values of \( r_{w } \)
together with their significance level \( p_{w } \), are listed in Table
1. The significance is established with 10000 sets of simulated fields each
with the same number of randomly distributed galaxies as the real fields.

\subsection{Is there an `identification bias'?}

From Table 1 we see that the VC sample, the one more similar to the PKS sample
used in BMG97, has the higher significance level, \( p_{w }=99.55\% \)
(the PKS itself has \( p_{w }=99.78\% \) using this test). However, the union 
of the VC, HJ and 1Jy samples, which includes only 8 more HJ quasars has a considerable 
lower value of \( r_{w } \) and of \( p_{w } \). If we analyze separately
these 8 ``non VC'' HJ quasars we see that they have \( r_{w }=0.97132 \)
and \( p_{w }=6.72\% \), i.e. they are strongly \textit{anticorrelated}
with foreground galaxies at the \( 93.28\% \) level. However, if we look at
the other seven HJ faint objects which have similar optical and radio fluxes
(\( B_{J}>18.6 \), \( S_{2.7}<0.8 \)) but were included in the VC catalog, they
have \( <r_{w }>=1.03805 \) and \( p_{w }=93.43\% \), that is, an
excess much stronger than that detected for the full HJ sample. If we put together
these 15 faint quasars, we get \( r_{w }=0.99953 \), that is, practically
no correlation whatsoever, positive or negative. Since magnitudes, radio-fluxes,
absolute luminosities, redshifts and galactic latitudes are very similar for
both faint ``minisamples'', the only remaining difference seems to be their
inclusion in the '96 version of the Veron-Cetty \& Veron catalog, something which 
only depends on the date when the quasars were identified. That happened in 1983 for the
7 VC quasars, whereas the ``non-VC'' redshifts were first published in
Drinkwater et al. (1997). If we compare the values of \( r_{w } \) of the
8 ``non-VC'' objects with the rest of quasars (86) from the HJ sample, a
Kolmogorov-Smirnov test shows that their \( r_{w } \) distributions are
incompatible at the \( 99.31\% \) level. This is rather puzzling, and barring
a statistical fluke, the only remaining possibility seems to be something which
might be called ``identification bias'': apparently the first radio sources
to be spectroscopically identified in a catalog tend to be those in regions of 
higher galaxy density. It is possible that due to the positional uncertainty of the
radio identifications, there was a tendency to start the identification
of a radio catalog with those fields which have more ``candidates'' close
to the radio position. Although in our case the samples are small, the differences
among them are so significant that one must conclude that any results about 
quasar-galaxy associations obtained with incomplete catalogs should be considered with 
great caution. 

The results of BMG97 are a good example: if we look at the 86 quasars in 
the HJ sample considered in that paper (where the 8 ``non-VC'' objects were not
included) one obtains \( r_{w }=1.01554 \) and \( p_{w }=99.14\% \).
The total HJ sample gives \( r_{w }=1.01178 \) and \( p_{w }=97.33\% \),
as we see from Table 1. Although the galaxy excess does not disappear
when we use the complete sample, the values of \( r_{w } \) and \( p_{w } \)
drop appreciably. If this happens with a sample which was almost \( (86/94=91\%) \)
complete, results as those of Tyson (1986) , Hintzen et al. (1991), or 
Burbidge et al. (1990), which use QSO samples extracted from incomplete compilations
as those of Hewitt \& Burbidge (1980) and the Veron-Cetty \& Veron (1984) catalogs, may 
be seriously biased, explaining the extremely strong amplitudes of \( w _{qg} \) 
found by these authors.

Therefore, it seems clear that valid samples for quasar-galaxy correlation measurements
should be complete or randomly extracted (e.g., depending on the celestial
coordinates) from complete catalogs, as the 1Jy and HJ. Note that the 1Jy has the highest value
of \( r_{w } \) for the samples in Table 1, i.e. it is the most strongly
correlated, in qualitative agreement with the magnification bias effect, 
although \( p_{w } \) is not very high due to the smaller size of the sample, 
which increases the statistical uncertainty.

\section{Comparison with theory}

To compare with theory, we shall assume that the biasing factor 
\( b=\delta _{g}/\delta _{m} \) is approximately constant within the relevant angular range. 
Two estimates of \( w_{qg} \) will be considered, that 
of Sanz, Martinez-Gonzalez \& Ben\'\i tez (1997), which takes into account the
nonlinear evolution of the power spectrum using the ansatz of Peacock \& Dodds (1996),  
and the more recent estimation of Ben\'\i tez \& Sanz (1999).

\subsection{Power spectrum-based estimate}

\label{power}

\begin{figure}
%\vspace{8.89cm}
\psfig{figure=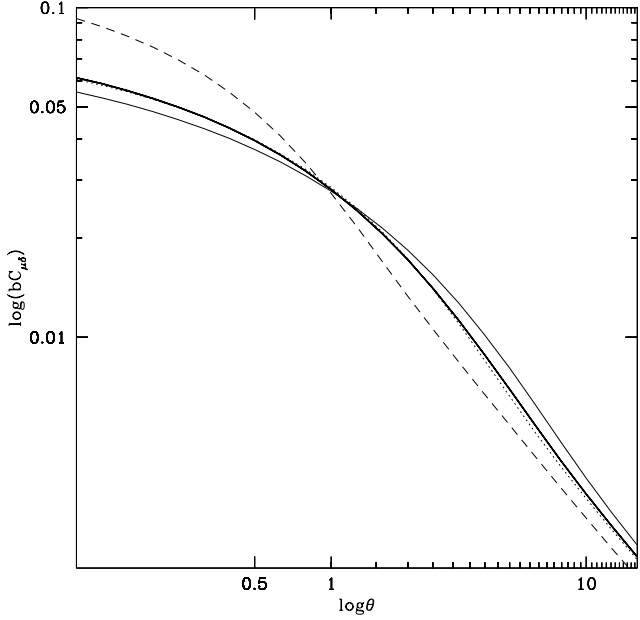,height=3.5in}
\caption{The product \protect\protect\protect\( bC_{\mu \delta }\protect \protect \protect \)
for several omega values, \protect\protect\protect\( \Omega_m =0.1\protect \protect \protect \)(dashed), 
\protect\protect\protect\( 0.4\protect \protect \protect \) (dotted) ,\protect\protect\protect\( 1\protect \protect \protect \)
(solid thin line). The thick line is \protect\protect\protect\( <bC_{\mu \delta }>\protect \protect \protect \). }
\end{figure}

In Sanz, Mart\'\i nez-Gonz\'alez \& Ben\'\i tez (1997) we showed that the the value of \( C_{\mu \delta } \) in Eq. (\ref{uno})
practically does not change if, instead of using the real galaxy and quasar distributions
for the calculations, it was assumed that all the galaxies and quasars were
respectively placed at redshifts \( z_{b} \) and \( z_{f} \). Our calculations
are normalized to the cluster abundance (Viana \& Liddle 1996) 
\( \sigma _{8}=0.6\Omega ^{-0.34+0.28\Omega -0.13\Omega ^{2}} \) 

For consistency we have to normalize the large-scale biasing factor 
as \( b=\sigma _{8}^{-1} \).
As we saw above, \( \omega _{qg}\propto bC_{\mu \delta } \), so this product
contains all the information relevant to the dependence of \( \omega _{qg} \)
on \( \Omega  \). The variation of \( b \) and of the amplitude of \( C_{\mu \delta } \)
with \( \Omega  \) somehow cancel and the product \( bC_{\mu \delta } \) depends
weakly on \( \Omega  \), as it is shown in Fig 5, where we plot \( bC_{\mu \delta } \)
calculated using delta-functions for the galaxy and quasar distributions with
\( z_{g}=0.15,z_{q}=1.4 \). We see that there are two different regimes for
\( w _{qg} \), one at small scales \( \theta \leq 1\arcmin  \) and the
other at larger scales, \( \theta >2\arcmin  \).

\begin{figure}
%\vspace{8.89cm}
\psfig{figure=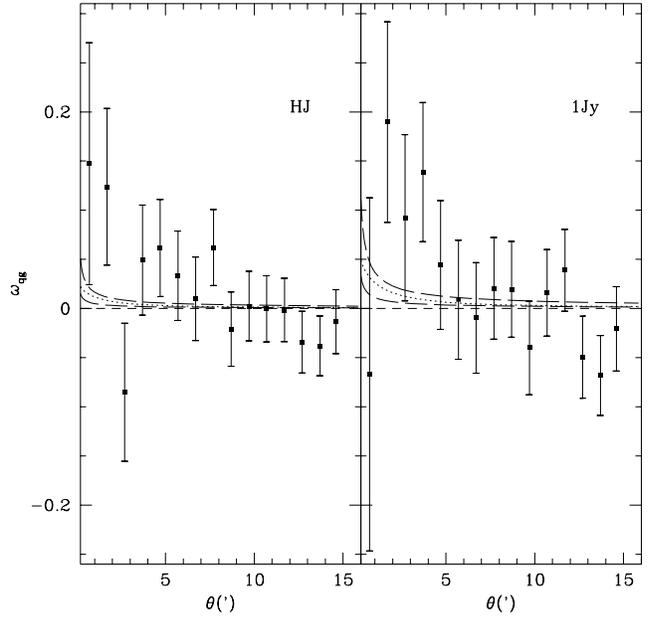,height=3.5in}
\caption{
Quasar-galaxy correlation for the HJ (to the left) and 1Jy 
(to the right) samples. The error bars are poissonian and the bin width is approximately
\protect\protect\protect\( 1\arcmin \protect \protect \protect \). The dotted
line corresponds to 
\protect\protect\protect\( (\alpha _{e}-1)<bC_{\mu }\delta >\protect \protect \protect \)
(see sec.\ref{power}). The two dashed lines correspond to the estimation of 
\protect\( w_{qg}\protect \)
of Sec. \ref{free} with 
\protect\( \Omega /b=0.3\protect \) (lower) and \protect\( \Omega /b=1\protect \)
(upper). We can see that in both cases the observed correlation is much larger
than the expected one.}
\end{figure}

The dependence of \( w _{qg} \) on \( \Omega  \) at small scales can
be approximated as 
\[
w _{qg}\approx A_{s}(\Omega)\log \theta +B_{s}\]
 This fit is valid for the angular scales \( 0.1\arcmin <\theta <1\arcmin  \). 
\( A_{s}(\Omega)=-0.029\Omega ^{0.37\log \Omega } \)
and \( B_{s} \) (which is equal to the value of \( w _{qg} \) at \( 1\arcmin  \))
is practically a constant, \( B_{s}=0.029\pm 0.02 \).  \( A_{s} \) quickly drops 
as \( \Omega  \) grows, although the amplitude of the variation
is relatively small, \( A_{s}=2.34 \) at \( \Omega =0.1 \) and \( A_{s}=1 \) for a flat
universe. At larger scales, \( \theta >>1\arcmin  \), \( w _{qg} \) is
reasonably well approximated by a law of the form 
\[
w _{qg}\approx A_{l}\theta ^{-1}\]
 The amplitude \( A_{l} \) also weakly depends on \( \Omega  \), with \( A_{l}=0.029\Omega ^{0.2} \),
but unlike the situation on small scales, it grows with the value of \( \Omega  \).

Since the variation of \( w_{qg} \) with \( \Omega  \) is small, the 
observations will only be compared with the product \( (\alpha _{e}-1)<bC_{\mu \delta }> \),
where \( <bC_{\mu \delta }>=\int ^{1}_{0}d\Omega bC_{\mu \delta }(\Omega ) \).
This comparison is shown in Fig 6a, where we plot \( w _{qg} \) as measured
from the HJ and 1Jy samples. It is clear that the observed correlation is much
stronger than the model prediction.

To estimate this mismatch we perform a maximum likelihood fit to the data of
the form 
\begin{equation}
w _{qg}=A(\alpha _{e}-1)<bC_{\mu \delta }>
\end{equation}
 and leave \( A \) as a free parameter. We found that for the HJ sample 
\( A=12.7\pm 6.5 \),
where the error limits enclose a \( 68\% \) confidence interval, and \( A>1 \)
at the \( 98.40\% \) level. The 1Jy sample also displays a much stronger correlation
than expected: \( A=10.2\pm 4.5 \), and \( A>1 \) at the \( 99.08\% \) level.
These results are not appreciably changed by varying 
\( bC_{\mu \delta }(\Omega ) \)
within the range \( 0.1<\Omega <1 \).

\subsection{Model independent estimate \label{free}}

Ben\'\i tez \& Sanz (1999) propose a model-free estimation of \( w_{qg} \), summarized
in Eq. (\ref{dos}). We can only apply it approximately, since it assumes
that the foreground and background galaxies are localized in 'thin' slices,
with the angular cosmological distance practically constant. This is
not strictly the case here, but a rough estimate can be found as \( w_{qg}\propto w_{gg} \)
and \( A_{qg}\approx A_{gg}Q(z_{f},z_{b})\Delta z_{f}(\alpha -1)\frac{\Omega }{b} \),
where for \( z_{b} \) and \( z_{f} \) we will take the median values of the
QSO (\( z_{b}=1.4 \)) and galaxy populations (\( z_{f}=0.15 \)) respectively.
Also for the galaxies \( \Delta z_{f}\approx 0.2 \), \( A_{gg}\approx 0.44 \). 
Since the proportionality factor is \( Q(0.15,1.4)\approx 0.45 \)
(Ben\'\i tez \& Sanz 1999), and we expect to measure a correlation amplitude
of \( \frac{A_{qg}}{\alpha -1}\approx 0.04\frac{\Omega }{b} \). A maximum likelihood
fit to the data using a function of the form \( w_{qg}\propto A\theta ^{-\gamma } \)
, taking a fixed \( \gamma =0.688 \) and leaving \( A \) as a free
parameter, yields \( \frac{A_{HJ}}{\alpha _{HJ}-1}=0.39\pm 0.20 \) and \( \frac{A_{1Jy}}{\alpha _{1Jy}-1}=0.31\pm 0.14 \), again much higher than expected. Given 
the redshifts of the background and foreground populations, including a 
cosmological constant \( \Lambda  \) would not change significantly the results.

\section{Discussion}

Although the observed results seem to qualitatively agree with the
lensing hypothesis (for instance, the amplitude of $w_{qg}$ divided by
$\alpha_e-1$ is almost constant for both the 1Jy and the HJ catalogs),
they cannot be quantitatively reconciled with it. For instance, from the results of 
the above paragraph and the estimate of Ben\'\i tez \& Sanz (1999) one would need 
\( \frac{\Omega }{b}\sim 10 \). There seems to be plenty of evidence for 
\( \Omega \simeq 1/3 \) (see e.g. Bahcall et al. 1999), and therefore the above ratio 
would imply absurdly low values of the biasing parameter $b$. In spite of some hints 
of `antibiasing' in low $\Omega$ LSS simulations (Jenkins et al. 1998), most 
measurements point at values of $b\approx 1$ for the APM galaxies 
(e.g. Frieman \& Gazta\~naga 1999), although on scales 
several times larger than the ones considered here ($R<1.5h^{-1}$ Mpc 
at $z<0.15$). 
  
It seems difficult to believe that the weak lensing theory is wrong by 
an order of magnitude. For instance, the calculation by Williams (1999)
taking into account second-order effects only changes in $\approx 10\%$ the 
first-order results. Are there 
other possible explanations for these high values of the QSO-galaxy correlation?

One option often considered is absorption by Galactic dust 
(Norman \& Williams 1999). It is easy to see that this effect would 
lead to a positive correlation between galaxies and high-z quasars: 
Let's suppose that \( \delta \tau (\theta ) \) represents a (small) 
fluctuation of the galactic extinction in the sky. The induced 
fluctuations in the galaxy and quasar number density, \( \delta n_{g} \) 
and \( \delta n_{q} \) respectively would be 
\( \delta n_{q}\approx -\alpha _{q}\delta \tau  \) and 
\( \delta n_{g}\approx -\alpha _{g}\delta \tau  \), where \( \alpha _{q} \) 
and \( \alpha _{g} \) are the slopes of the optical number counts magnitude 
distributions of quasars and galaxies. Therefore, 
\begin{equation}
w^{\tau }_{qg}=<\delta n_{q}\delta n_{g}>\approx \alpha _{q}\alpha _{g}C_{\tau \tau }
\end{equation} 
where \( C_{\tau\tau} \) 
is the dust-dust correlation function in e.g. 
Schlegel, Finkbeiner, \& Davis (1998). Although the quasar-galaxy 
correlation \( w ^{\tau }_{qg} \) also increases with 
the value of the slope \( \alpha _{q} \), there are significant 
differences with the correlation generated by the magnification bias effect: 
\( w ^{\tau }_{qg} \) is always positive, even for values of \( \alpha _{q} \)
smaller than \( 1 \). This could allow to distinguish between the contribution
of each effect to the total correlation. The best observational results on 
\( C_{\tau \tau } \) available so far, those of Finkbeiner, Schlegel \& Davis (1998) 
do not resolve adequately the angular scales on which \( w_{qg} \) is measured 
in this paper. A tentative extrapolation of the \( C_{\tau \tau } \) presented
in the above mentioned paper is too flat to explain the measured correlations. 
Future measurements of $C_{\tau\tau}$ on small scales will serve to establish the 
contribution of $w^\tau_{gg}$ to the total, observed $w_{gg}$.

It is also interesting to explore the contribution to $w_{qg}$ on 
relatively large scales, $> 1'$,  from strong lensing effects. 
Let's suppose that a fraction $f_S$ of the quasars in a sample is 
strongly lensed (not necessarily multiply imaged) by individual 
foreground galaxies. These galaxies, which will be very close to the quasar 
positions, are correlated with other galaxies, following a correlation 
function $w_{gg}$. If the typical quasar-lens distance is considerably 
smaller that the galaxy-galaxy correlation scale, this would indirectly 
cause a ``strong-lensing'' quasar--galaxy correlation $w_{qg}^S\sim f_Sw_{gg}$. 
Since the ``weak lensing'' correlation estimated above is 
$w^W_{qg}\approx 0.02-0.04w_{gg}$, relatively small values of $f_S$ 
could generate a correlation $w^S_{qg}$ of comparable amplitude. 
A quick inspection of the Digital Sky Survey images around the quasars 
in the HJ sample reveals several cases of very close, $\theta<$ few 
arcsec `associations' in a sample of $\sim 100$ quasars. Unfortunately, 
and due to the poor quality of these images (some of the `associated' 
objects may be stars or defects), it is difficult to give an exact value 
of $f_S$ for this sample. However, if an appreciable fraction of these 
objects are galaxies, this effect would contribute significantly to the 
total amplitude of $w_{qg}$. Obviously this point merits a more detailed 
consideration with improved observations.

Last but not least, another possibility is that the observed correlation $w_{qg}$ 
is affected by unforeseen systematic effects, or due to a statistical fluke. 
This sounds very unlikely given the variety of quasar and galaxy catalogs 
considered in $w_{qg}$ studies (Ben\'\i tez 1997). But it should not be 
forgotten that many of the positive results reported in the literature are based 
on bright, radio--loud quasars, and they are so scarce in the sky that overlaps among 
the samples are unavoidable.

As in most scientific controversies, the solution will eventually come 
from more and better data: ongoing surveys as the Sloan Digital Sky 
Survey (Gunn \& Weinberg 1995) or the 2dF (Folkes et al. 1999) will 
provide huge QSO and galaxy catalogs, from which it will be possible 
to select complete samples well suited to measure $w_{qg}$. 

\section{Conclusions}

Gravitational lensing predicts an enhancement of the density of background QSOs
around foreground galaxies. We measure this QSO---galaxy correlation \( w_{qg} \)
for two \textit{complete} samples of radio-loud quasars, the southern 1Jy and
Half-Jansky samples. The existence of a positive correlation between distant 
\( z\sim 1 \) quasars and \( z\approx 0.15 \) galaxies is confirmed at a \( p=99.0\% \)
significance level (\( >99.9\% \) if previous measurements on the northern hemisphere 
are included). A comparison with the results obtained for incomplete quasar catalogs 
(e.g. the Veron-Cetty and Veron compilation) suggests the existence of an 
`identification bias', which spuriously increases the estimated amplitude 
of \( w _{qg} \) for incomplete samples. This effect could explain most of 
the very strong quasar--galaxy statistical associations found in the literature. 
Nevertheless, the value of \( w_{qg} \) that we measure in our complete catalogs 
is still considerably higher than the predictions from weak gravitational lensing theory.
Including the effects of strong lensing could help to explain this discrepancy.

\section*{Acknowledgements} 
We thank D. Finkbeiner for kindly providing the numerical 
values of $C_{\tau\tau}$ and for helpful comments.


\begin{thebibliography}{99}

\bibitem[Bahcall et al. 1999]
{1999Sci...284.1481B} Bahcall, N. A., Ostriker, J. P., Perlmutter, S. 
\& Steinhardt, P. J. 1999, Science, 284, 1481 

\bibitem[]{}Bartelmann, M. 1995, A\&A, 298, 661

\bibitem[Bartelmann \& Schneider 1993a]{bar93} Bartelmann, M. 
\& Schneider, P. 1993a, A\&A, 268, 1 

\bibitem[Bartelmann \& Schneider 1993b]{1993A&A...271..421B} Bartelmann, M. 
\& Schneider, P. 1993b, A\&A, 271, 421 

\bibitem[]{}Bartelmann, M., \& Schneider, P. 1994, A\&A, 284, 1

\bibitem[]{}Bartsch, A., Schneider, P. \& Bartelmann, M., 1997, A\&A, 319, 375

\bibitem[Ben\'\i tez 1997]{} Ben\'\i tez, N. 1997, PhD. thesis, Universidad de Cantabria

\bibitem[]{}Ben\'{\i}tez, N., \& Mart\'{\i}nez-Gonz\'{a}lez, 1995, (BMG95) ApJLett, 339, 53

\bibitem[]{}Ben\'{\i}tez, N., \& Mart\'{\i}nez-Gonz\'{a}lez, 1997, (BMG97) ApJ, 477, 27

\bibitem[]{}Ben\'{\i}tez, N., Mart\'{\i}nez-Gonz\'{a}lez, E., Gonz\'{a}lez-Serrano,
J.I., \& Cay\'{o}n L. 1995, AJ, 109, 935

\bibitem[Ben\'\i tez, Mart\'\i nez-Gonzalez \& Mart\'in-Mirones 
1997]{1997A&A...321L...1B} Ben\'\i tez, N., Mart\'\i nez-Gonz\'alez, E. \& 
Mart\'\i n-Mirones, J. M. 1997, A\&A, 321, L1 

\bibitem[Ben\'\i tez \& Sanz 1999]{1999ApJ...525L...1B} Ben\'\i tez, 
N. \& Sanz, J. L. 1999, ApJLett, 525, L1 

\bibitem[]{}Borgeest, U., von Linde, J., Refsdal, S. 1991, A\&A, 251, L35

\bibitem[]{}Boyle, B.J., Fong, R. \& Shanks, T. 1988, MNRAS, 231, 897

\bibitem[Burbidge Hewitt Narlikar \& Gupta 1990]{1990ApJS...74..675B} 
Burbidge, G., Hewitt, A., Narlikar, J. V. \& Gupta, P. D.  1990, ApJS, 74, 
675 

\bibitem[Canizares 1981]{1981Natur.291..620C} Canizares, C. R. 1981, Nature, 
291, 620 

\bibitem[Crampton, Cowley \& Hartwick 1989]{cramp} Crampton, 
D. , Cowley, A. P. \& Hartwick, F. D. A. 1989, ApJ, 345, 59 

\bibitem[Croom \& Shanks 1999]{1999MNRAS.307L..17C} Croom, S. M. \& Shanks, 
T. 1999, MNRAS, 307, L17 

\bibitem[Dolag \& Bartelmann 1997]{1997MNRAS.291..446D} Dolag, K.  \& 
Bartelmann, M.  1997, MNRAS, 291, 446 

\bibitem[Drinkwater, Webster, Thomas \& Millar 1992]{1992PASAu..10....8D} 
Drinkwater, M. J., Webster, R. L., Thomas, P. A. \& Millar, E. 1992, 
Proceedings of the Astronomical Society of Australia, 10, 8 

\bibitem[Drinkwater et al. 1997]{1997MNRAS.284...85D} Drinkwater, M. J., et 
al. 1997, MNRAS, 284, 85 

\bibitem[Ferreras Benitez \& Martinez-Gonzalez 1997]{1997AJ....114.1728F} 
Ferreras, I. , Ben\'\i tez, N.  \& Mart\'\i nez-Gonz\'alez, E.  1997, AJ, 114, 1728 

\bibitem[Finkbeiner, Schlegel \& Davis 1998]{doug} 
Finkbeiner, D. P., Schlegel, D. J. \& Davis, M. 1998, Berlin Springer 
Verlag Lecture Notes in Physics, v.506, 506, 367 

\bibitem[Folkes et al. 1999]{1999MNRAS.308..459F} Folkes, S. , et al. 1999, 
MNRAS, 308, 459 

\bibitem[Frieman \& Gazta\~naga 1999]{1999ApJ...521L..83F} Frieman, J. A. \& 
Gazta\~naga, E.  1999, ApJLett, 521, L83 

\bibitem[Fugmann 1988]{1988A&A...204...73F} Fugmann, W. 1988, A\&A, 204, 73 

\bibitem[]{}Fugmann, W. 1990, A\&A, 240, 11

\bibitem[Gunn \& Weinberg 1995]{sdss}Gunn, J. E., \& Weinberg, D. H. 1995, 
in Wide-Field Spectroscopy and the Distant Universe, 
ed. S. J. Maddox \& A. Arag\'on-Salamanca (Singapore: World Scientific)

\bibitem[Hammer \& Le Fevre 1990]{1990ApJ...357...38H} Hammer, F. \& Le 
Fevre, O. 1990, ApJ, 357, 38 

\bibitem[]{}Hartwick, F.D.A \& Schade, D. 1990, ARAA, 28, 437

\bibitem[Hawkins \& Veron 1995]{hawk} Hawkins, M. R. S. \& 
Veron, P. 1995, MNRAS, 275, 1102 

\bibitem[Hewett, Foltz & Chafee 1995]{lbqs}Hewett, P.C., Foltz, C.B.\& Chafee, F.H., 1995, AJ, 109, 1498

\bibitem[Hewitt \& Burbidge 1980]{hb} Hewitt, A. \& 
Burbidge, G. 1980, ApJS, 43, 57 

\bibitem[]{}Hintzen, P., Romanishin, W., \& Vald\'{e}s, F., 1991, ApJ, 371, 49

\bibitem[Jenkins et al. 1998]{1998ApJ...499...20J} Jenkins, A., et al. 
1998, ApJ, 499, 20 

\bibitem[Keeton Kochanek \& Falco 1998]{1998ApJ...509..561K} Keeton, C. R., 
Kochanek, C. S. \& Falco, E. E. 1998, ApJ, 509, 561 

\bibitem[Maddox et al. 1990]{apm} 
Maddox, S. J., Efstathiou, G., Sutherland, W. J. \& Loveday, J. 1990, 
MNRAS, 242, 43P 

\bibitem[]{}Maoz, D., 1995, ApJ, 455, 115

\bibitem[Narayan 1989]{1989ApJ...339L..53N} Narayan, R.  1989, ApJLett, 339, 
L53 

\bibitem[Norman \& Impey 1999]{1999AJ....118..613N} Norman, D. J. \& Impey, 
C. D. 1999, AJ, 118, 613 

\bibitem[Norman \& Williams 1999]{norman}Norman, D.J. \& Williams, L.L.R. 1999, 
astro-ph/9908177 

\bibitem[Peacock 1982]{1982MNRAS.199..987P} Peacock, J. A. 1982, MNRAS, 
199, 987 

\bibitem[Peacock \& Dodds 1996]{1996MNRAS.280L..19P} Peacock, J. A. \& 
Dodds, S. J. 1996, MNRAS, 280, L19 


\bibitem[Rodrigues-Williams \& Hogan 1994]{1994AJ....107..451R} 
Rodrigues-Williams, L. L. \& Hogan, C. J. 1994, AJ, 107, 451 

\bibitem[Romani \& Maoz 1992]{1992ApJ...386...36R} Romani, R. W. \& Maoz, 
D.  1992, ApJ, 386, 36 

\bibitem[Sanz, Mart\'\i nez-Gonz\'alez \& Ben\'\i tez 1997]{1997MNRAS.291..418S} Sanz, J. L., 
Mart\'\i nez-Gonz\'alez, E. \& Ben\'\i tez, N. 1997, 
MNRAS, 291, 418 

\bibitem[Schlegel Finkbeiner \& Davis 1998]{1998ApJ...500..525S} Schlegel, 
D. J., Finkbeiner, D. P. \& Davis, M.  1998, ApJ, 500, 525 

\bibitem[]{}Schneider, P., Ehlers, J., \& Falco, E.E. 1992, Gravitational Lenses (Heidelberg:
Springer)

\bibitem[Seitz \& Schneider 1995]{1995A&A...302....9S} Seitz, S. \& 
Schneider, P. 1995, A\&A, 302, 9 

\bibitem[]{}Smith, R.J., Boyle, B.J.\& Maddox, S.J., preprint astro-ph/9506093

\bibitem[Stickel, Meisenheimer & K\"{u}hr 1994]{sti94}
Stickel, M., Meisenheimer, K., \& K\"{u}hr, H. 1994, A\&AS, 105, 211

\bibitem[Stickel et al. 1996]{sti96}
Stickel, M., Rieke, M.J., Rieke, G.H. \& K\"{u}hr, H. 1996, A\&A, 306, 49

\bibitem[]{}Thomas, P.A., Webster, R.L., \& Drinkwater, M.J. 1994, MNRAS, 273, 1069

\bibitem[]{}Tyson, J.A., 1986, AJ, 92, 691

\bibitem[]{}Webster, R.L., Hewett, P.C., Harding, M.E., \& Wegner, G.A. 1988, Nature, 336,
358

\bibitem[Williams \& Irwin 1998]{1998MNRAS.298..378W} Williams, L. L. R. \& 
Irwin, M.  1998, MNRAS, 298, 378 

\bibitem[]{}Williams, L.L.R. 1999, submitted to ApJ, astro-ph/9908318

\bibitem[]{}Wright, A. \& Otrupcek, R.E. PKSCAT90: Radio Source Catalogue and Sky Atlas
( Australia Telescope National Facility, Epping, NSW, 1990)

\bibitem[Veron-Cetty \& Veron 1984]{ver84} Veron-Cetty, M. 
-P. \& Veron, P. 1984, ESO Scientific Report, Garching: European Southern 
Observatory (ESO), 1984,  

\bibitem[Veron-Cetty \& Veron 1996]
{ver96}Veron-Cetty, M.P. \& Veron, P., 1996, 
European Southern Observatory Scientific Report, 17, 1 

\bibitem[Viana and Liddle 1996]{1996MNRAS.281..323V} 
Viana, P. T. P. and Liddle, A. R. 1996, MNRAS, 281, 323 

\bibitem[Yentis et al. 1992]
{yen92}
Yentis, D.J, Cruddace, R.G., Gursky, H., Stuart, B.V., Wallin, J.F., Mac-Gillivray,
H.T. \& Thompson, E.B., 1992, Digitised Optical Sky Surveys, Ed. H.T. Mac-Gillivray
and E.B. Thompson, Kluwer,Dordrecht, 67

\end{thebibliography}
\end{document}